\documentclass[aps,twocolumn,preprintnumbers,amsmath,amssymb,superscriptaddress]{revtex4-2}

\usepackage[T1]{fontenc}
\usepackage[latin9]{inputenc}
\usepackage{color}
\usepackage{amsmath}
\usepackage{amssymb}
\usepackage{graphicx}
\usepackage{esint}

\makeatletter

\usepackage{amssymb}
\usepackage{amsmath}
\usepackage{graphicx}
\usepackage[colorlinks=true,linkcolor=blue,anchorcolor=red,citecolor=blue,urlcolor=blue]{hyperref}
\usepackage[caption=false]{subfig}
\usepackage{lipsum}

\usepackage{soul}

\makeatother

\DeclareMathOperator{\link}{Link}

\usepackage{babel}
\begin{document}
	
\renewcommand{\figurename}{Fig.}

\title{Hermitian topologies originating from non-Hermitian braidings}
 
\author{W.\ B. Rui}
\email{wbrui@hku.hk}
\address{Department of Physics and HKU-UCAS Joint Institute for Theoretical
and Computational Physics at Hong Kong, The University of Hong Kong,
Pokfulam Road, Hong Kong, China}

\address{HK Institute of Quantum Science \& Technology, The University of Hong Kong,
	Pokfulam Road, Hong Kong, China}

\author{Y.\ X. Zhao}
\email{yuxinphy@hku.hk}
\address{Department of Physics and HKU-UCAS Joint Institute for Theoretical
	and Computational Physics at Hong Kong, The University of Hong Kong,
	Pokfulam Road, Hong Kong, China}

\address{HK Institute of Quantum Science \& Technology, The University of Hong Kong,
	Pokfulam Road, Hong Kong, China}

\author{Z.\ D. Wang}
\email{zwang@hku.hk}
\address{Department of Physics and HKU-UCAS Joint Institute for Theoretical
	and Computational Physics at Hong Kong, The University of Hong Kong,
	Pokfulam Road, Hong Kong, China}

\address{HK Institute of Quantum Science \& Technology, The University of Hong Kong,
	Pokfulam Road, Hong Kong, China}


\begin{abstract}
The complex energy bands of non-Hermitian systems braid in momentum space even in one dimension. Here, we reveal that the non-Hermitian braiding underlies the Hermitian topological physics with chiral symmetry under a general framework that unifies Hermitian and non-Hermitian systems. Particularly, we derive an elegant identity that equates the linking number between the knots of braiding non-Hermitian bands and the zero-energy loop to the topological invariant of chiral-symmetric topological phases in one dimension. Moreover, we find an exotic class of phase transitions arising from the critical point transforming different knot structures of the non-Hermitian braiding, which are not included in the conventional Hermitian topological phase transition theory. 
Nevertheless, we show the bulk-boundary correspondence between the bulk non-Hermitian braiding and boundary zero-modes of the Hermitian topological insulators. Finally, we construct typical topological phases with non-Hermitian braidings, which can be readily realized by artificial crystals.
\end{abstract}

\maketitle

\section{INTRODUCTION}

The concept of topology captures properties unchanged under continuous deformations. It can be pictorially illustrated by knots formed by braids, which cannot be unknotted by continuous stretching. While initially introduced to describe the behavior of anyons~\cite{leinaas_theory_1977,Wilczek_1982,Nayak_rmp}, braiding was recently actively investigated in momentum-space band structures~\cite{wu_non-abelian_2019,Yang_nodal_line_braiding,wang_intrinsic_2021,park_non-abelian_2021,bouhon_non-abelian_2020,guo_experimental_2021,jiang_experimental_2021,peng_phonons_2022,Yang_nh_knots,tang_exceptional_2020,Wang_nh-EPs,tang_experimental_2021,deng_hopf_2013,tomma_nodal_line_2017,yang_nodal_line_2018,zhao_nodal_line_2020}, where braids appear in various scenes, such as nodal lines in topological semimetals~\cite{wu_non-abelian_2019,Yang_nodal_line_braiding,wang_intrinsic_2021,park_non-abelian_2021}, the trajectories of Weyl or Dirac points in adiabatic time evolution~\cite{bouhon_non-abelian_2020,guo_experimental_2021,jiang_experimental_2021,peng_phonons_2022}, and exceptional rings in non-Hermitian systems~\cite{Yang_nh_knots,tang_exceptional_2020,Wang_nh-EPs,tang_experimental_2021}.

Another active trend in recent research is to generalize elements of Hermitian topological phases to non-Hermitian systems~\cite{Esaki_PRB_2011,Leykam_PRL_2017,xiong_why_2018,Menke_PRB_2017,Xu_PRL_2017,Lee_PRL_2016,shen_PRL_2018,zhoubulkarc018,ruiIntertwined2022,zhang_PRL_2020,gongNonHermitian2018,ruiPT2019a,kawabataSymmetry2019a,Gong_PRB_2019,bergholtzExceptional2021}. Fundamental topological concepts, such as topological invariants, bulk-boundary correspondence, and topological stability, were reconsidered by incorporating non-Hermitian features~\cite{gongNonHermitian2018,kawabataSymmetry2019a,Gong_PRB_2019,bergholtzExceptional2021,berry_physics_2004,heiss_physics_2012,miri_exceptional_2019,katoPerturbationTheor1995,Kawabata_ep_2019}, such as gain-and-loss processes and exceptional points. This leads to novel topological classifications, and a plethora of non-Hermitian topological phase with exotic non-Hermitian topological phenomena~\cite{gongNonHermitian2018,kawabataSymmetry2019a,bergholtzExceptional2021,berry_physics_2004,heiss_physics_2012,miri_exceptional_2019,katoPerturbationTheor1995,Kawabata_ep_2019,ruiExceptional2019,bernard2002,zhounhsymmetry2019,lieuSymmetryClasses2018,budichSymmetry2019,ruiSpatial2021,leeAnatomySkin2019,Gong_PRB_2019,kunstBiorthogonal2018,wang_skin_2018,Murakami_prl_2019,Fang_winding_2020,Masatoshi_winding_2020,xiao_non-hermitian_2020,helbig_generalized_2020}.  Particularly, massive number of states reside on the boundary in the non-Hermitian skin effect~\cite{kunstBiorthogonal2018,wang_skin_2018,BorgniaPRL2020,Fang_winding_2020,Masatoshi_winding_2020,Murakami_prl_2019,xiao_non-hermitian_2020,helbig_generalized_2020}, rather than a few boundary states for Hermitian topological insulators.

Non-Hermitian physics and braiding are not independent. The energy spectra of non-Hermitian systems are complex numbers, and hence the complex bands can naturally braid in the complex plane even through a 1D Brillouin zone. This leads to a recent classification of non-Hermitian 1D systems by the knot structures~\cite{liHomotopical2021,wojcikHomotopy2020,Zhao2021prl,huKnotEP2021,wangBraiding2021}.

In this paper, we reverse the conventional reasoning by understanding Hermitian topological physics in terms of non-Hermitian braiding in one dimension. \textit{A priori}, there is a formal one-to-one map between a Hermitian Hamiltonian $H$ in class AIII and a non-Hermitian operator $Q$ in the basis specified by the eigenstates of the chiral symmetry. Using the map, we find that a number of fundamentals of Hermitian topology acquire more pictorial understandings from the braiding structure of the non-Hermitian bands. Particularly, we rigorously show that the abstractly defined topological invariant of the Su-Schrieffer-Heeger (SSH) model can be visualized as the linking number between the knots of the complex non-Hermitian bands and the zero-energy loop. Since the left and right zero-modes of $Q$ are exactly the zero modes of $H$ with opposite chirality, we see that the end zero modes of the Hermitian insulator can be regarded as consequences of the bulk non-Hermitian braiding. Nevertheless, the braiding rationale can give rise to a novel class of topological phase transitions characterized by the critical points where the knot structure of the non-Hermitian bands transforms, which go beyond the conventional theory of topological phase transitions. Moreover, we construct typical topological phases using our theory, which can be readily realized by artificial crystals.

\section{Framework}

\begin{figure*}
	\includegraphics[width=0.9\textwidth]{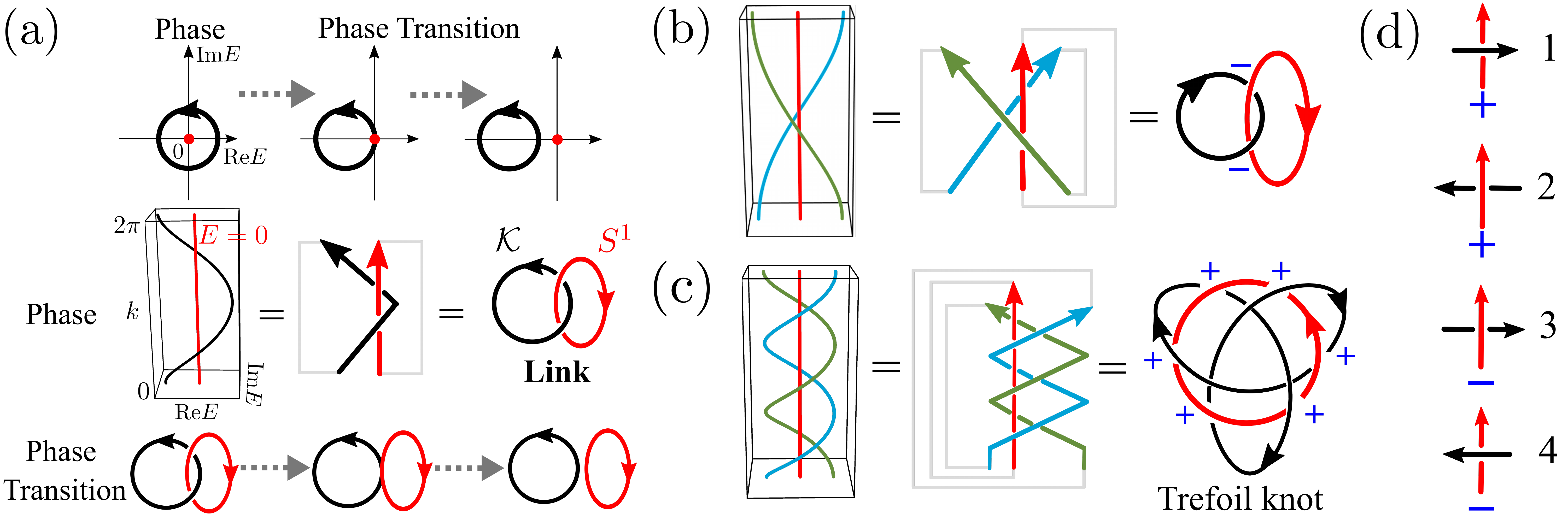}
	\caption{Non-Hermitian braidings as the origin of Hermitian topological phases. (a) Upper panel: the
		conventional 2D picture of topological phase and its transition. Middle panel: The single band braids trivially and forms a knot of circle. The 3D picture shows the topological origin as the linking between knot (black) and zero-energy loop (red). Lower panel: 3D picture of unlinking as the phase transition. (b) Knot of circle formed by the braiding of bands. (c) Trefoil knot formed by 
		the braiding of bands. (d) Illustrations of four types of crossings in the algorithm to calculate the
		topological invariant.
		\label{fig:link}}
\end{figure*}

Let us start with a non-Hermitian system described by the Hamiltonian
$Q$. Using $Q$, a Hermitian Hamiltonian
$H$ with chiral symmetry $S$ can be built as 
\begin{equation}
H=\left(\begin{array}{cc}
0 & Q\\
Q^{\dagger} & 0
\end{array}\right),\quad S=\left(\begin{array}{cc}
\mathbb{I} & 0\\
0 & -\mathbb{I}
\end{array}\right),\quad SHS^\dagger=-H.\label{eq:chiral-hal}
\end{equation}
Here $\dagger$ denotes the conjugate transpose and 
$\mathbb{I}$ the identity matrix. For the eigeneqution 
$H|\Psi\rangle=E|\Psi\rangle$, the chiral symmetry 
results in another set of eigenmodes $\{-E,S|\Psi\rangle\}$. 
Thus, the energy spectrum is symmetric with respect 
to zero energy.

A key observation is that for zero energy, i.e., $E=0$, the 
chiral-symmetric Hermitian $H$ is decoupled to non-Hermitian 
$Q$ and $Q^\dagger$. Though seemingly minor, we will show that
the non-Hermitian topological properties of $Q$ with respect to 
this zero energy dictate the bulk topology, 
topological phase transition, and bulk-boundary correspondence of the
Hermitian $H$. Thus, the zero energy serves as a guideline in our study.

Since zero-energy eigenstates of $H$ are 
at the same time the eigenstates of $S$, they can 
be classified as left-handed (right-handed) according to the 
eigenvalues $+1$$(-1)$ of the chiral operator. Introduce 
$|\Psi\rangle=(\psi_{+},\psi_{-})^{T}$ as the eigenstates of 
$H$. The zero-energy eigenstates ($|\Psi_{\pm}\rangle$) of 
different chiralities are obtained as
\begin{equation}
H|\Psi_{\pm}\rangle=0:\quad|\Psi_{+}\rangle=(\psi_{+},0)^{T},|\Psi_{-}\rangle=(0,\psi_{-})^{T},\label{eq:chiral-zero}
\end{equation}
which satisfies $S|\Psi_{\pm}\rangle=\pm|\Psi_{\pm}\rangle$. Clearly,
$H$ shares the same eigenstates with $Q$ and $Q^{\dagger}$, as
\begin{equation}
E=0:\quad Q^{\dagger}\psi_{+}=0,\,\quad Q\psi_{-}=0.\label{eq:Q-zero}
\end{equation}
Thus, we may regard that the zero-energy modes in chiral-symmetric
Hermitian systems originate from the non-Hermitain components. In 
contrast, non-zero-energy eigenstates of $Q$ can not retain in $H$~\cite{footnote}.

 \vspace{5cm}

\section{Origin of Hermitian topological phases: braidings of non-Hermitian energy bands}

We consider $Q$ with no symmetry (non-Hermitian class A) and $H$ with only chiral symmetry (Hermitian class AIII) in Eq.~(\ref{eq:chiral-hal})~\cite{chiuClassification2016,kawabataSymmetry2019a}.
In one dimension (1D), both $H$ and $Q$ are characterized by the winding number of
\begin{equation}
	w=\frac{1}{2\pi i}\oint_{\text{BZ}^{1}}dk\cdot\nabla_{k}\log\det\mathcal{Q}(k), \label{eq:chiral-winding}
\end{equation}
where $k$ denotes the momentum and $\mathcal{Q}(k)$ the corresponding Hamiltonian in the 1D Brillouin zone ($\text{BZ}^{1}$)~\cite{SSH1988,asboth2016}. In this regard, $H(k)$ and $Q(k)$ share the same topology, and formerly, correspondences between the two have been established~\cite{royFloquet2017,gongNonHermitian2018,kawabataSymmetry2019a,leeHermitian2019}. It turns out the "one-to-one correspondence"~\cite{royFloquet2017,kawabataSymmetry2019a} describing the topological equivalence between the Hermitian topology of $H$ and the point-gap topology~\cite{pointgap} of $Q$ is quite useful, and a systematic classification of the point-gap topology of $Q$ using $H$ has been worked out in higher dimensions and in other symmetry classes.

It shall be noted that the "one-to-one correspondence" above is based on the point-gap topology of $Q$. Recently, however, it has been shown that the non-Hermitian $\mathcal{Q}(k)$ actually has a 3D braiding topology in the ($k$, Re$E$, Im$E$)-space~\cite{liHomotopical2021,wojcikHomotopy2020,Zhao2021prl,huKnotEP2021} as shown in Figs.~\ref{fig:link}(a,b,c), which goes beyond the point-gap topology.
In view of this, the well-established equivalence relation between $Q$ and $H$ may not be sufficient.

However, the topological invariant in Eq.~\eqref{eq:chiral-winding} is not able to capture 
all the physics, because
it is essentially a 2D quantity, while the braiding structure of $Q(k)$ lives in 3D. Thus, it is necessary 
to develop a 3D picture to understand the topological origin and further explore the relation between $Q$ and $H$. In the middle panel of Fig.~\ref{fig:link}(a), the single band of $\mathcal{Q}(k)$ braids trivially and 
forms a knot of circle (black) over BZ in 3D, while the zero-energy point
forms a closed loop (red). To have a non-trivial phase,
the knot ($\mathcal{K}$) and the zero-energy loop ($S^1$) must form a link. 
Note that the zero-energy loop $S^1$ is defined by a constant function $E(k)=0$, $\forall k\in [0,2\pi)$, which forms a
loop due the periodicity of BZ.
Then, the winding number in Eq.~\eqref{eq:chiral-winding} actually reflects the linking number between the two,
\begin{equation}
	w=\link(\mathcal{K},S^1). \label{eq:link}
\end{equation}
It is clear now that the topological phase transition, as shown in the lower panel of Fig.~\ref{fig:link}(a), is triggered by the unlinking process between $\mathcal{K}$ and $S^1$. This is in contrast with the conventional understanding of the phase transition induced by shifting the energy band away from the zero-energy point, as shown by the upper panel of Fig.~\ref{fig:link}(a).

The linking number of Eq.~\eqref{eq:link} goes beyond the previously mentioned correspondence between $\mathcal{H}(k)$ and $\mathcal{Q}(k)$, and can faithfully describe the topological origin of $\mathcal{H}(k)$.
This is more evident for general cases of multi-band $\mathcal{Q}(k)$. In these cases,
the $n$ separable bands of $\mathcal{Q}(k)$ braid with each other [e.g., Figs.~\ref{fig:link}(b, c)], and form knots due to
the periodicity of BZ. 
As different braiding structures may correspond to the same point-gap topology, 
the 2D winding number by Eq.~\eqref{eq:chiral-winding}, while valid in describing point-gap topology, may be blind to these 3D braiding structures. 
The 3D linking number, on the other hand,
can faithfully tell us that the topological phases of $\mathcal{H}(k)$ originate from the non-Hermitian braidings in $\mathcal{Q}(k)$.

Compared to Eq.~\eqref{eq:chiral-winding}, the linking number may also simplify the calculation~\cite{calnote}, because it can be obtained by just counting the crossings,
\begin{equation}
	\link(\mathcal{K},S^1)=\frac{n_1+n_2-n_3-n_4}{2},
\end{equation}
where $n_i$ denotes the number of four types of crossings, as shown in Fig.~\ref{fig:link}(d). For instance,
by simply counting the crossings, the winding number of the knot of circle in (b) and the Trefoil knot in (c) is obtained as
$w=-1$ and $+3$, respectively.

\begin{figure}[t]
	\includegraphics[width=0.9\columnwidth]{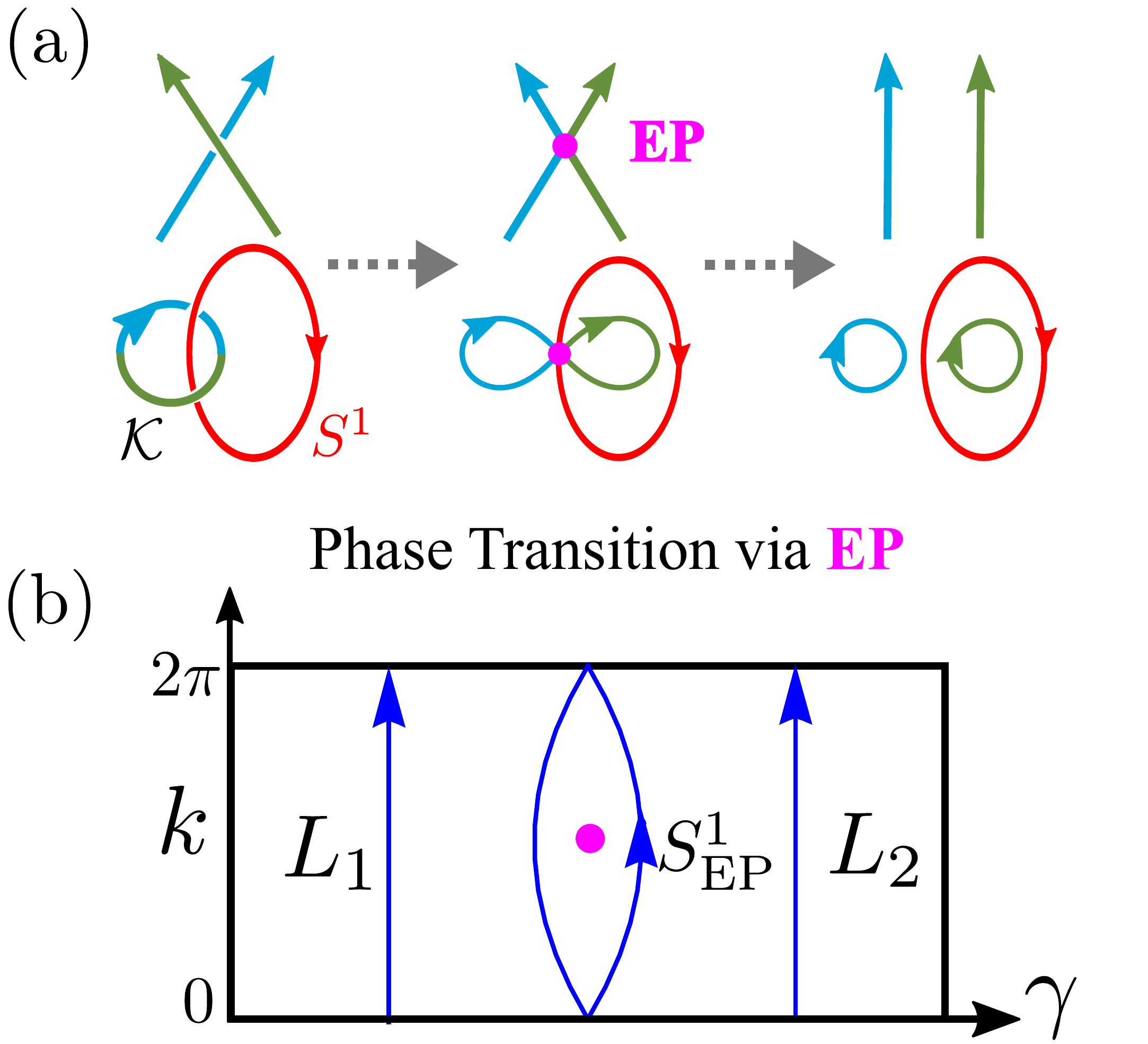}
	\caption{Phase transition via critical point in non-Hermitian braidings. (a) The transformation of non-Hermitian braiding in $\mathcal{Q}(k)$ via EP. (b) The transformation via EP can induce Hermitian topological phase transition in $\mathcal{H}(k)$.\label{fig:EP}}
\end{figure}

\vspace{0.3cm}

\section{Hermitian topological phase transition via critical point in non-Hermitian braidings}

The above discussion of the non-Hermitian braidings as the origin of Hermitian topology leads to our discovery of a different kind of topological phase 
transitions, which goes beyond the conventional theory of topological phase transitions shown in the first row of Fig.~\ref{fig:link}(a). 

As illustrated by Fig.~\ref{fig:EP}(a), where the two bands braid with each other, the linking number can also be changed by the transformation of the braiding structure. Here, the unlinking between $\mathcal{K}$ and $S^1$ is realized by the change of braiding of $\mathcal{Q}(k)$.
It happens via the critical point where two bands intersect on the zero-energy loop.
Such a critical point is 
a unique non-Hermitian degeneracy, namely, exceptional point (EP)~\cite{shen_PRL_2018,Zhao2021prl}, which results in the phase transition in $\mathcal{H}(k)$.

Suppose that the phase transition is induced by tuning an extra parameter $\gamma$. 
Then the momentum and this extra parameter forms a 2D $\left(k,\gamma\right)$-space, where EP could appear in this 2D space. As shown in Fig.~\ref{fig:EP}(b), $k_\text{EP}$ denotes the position of EP and $S_\text{EP}^1$ denotes the loop that encircles $k_\text{EP}$. To show that EPs can induce phase transitions in Hermitian systems, recall that for order-2 EPs in 2D space, 
there is an associated topological invariant of vorticity
~\cite{shen_PRL_2018,yangDoubling2021}
\begin{equation}\label{vorticity}
v_\pm=-\frac{1}{2\pi}\oint_{S_\text{EP}^1}d\mathbf{k}\cdot\nabla_{\mathbf{k} } \arg\left[E_+\left(\mathbf{k}\right){-E}_-\left(\mathbf{k}\right)\right],
\end{equation}
where $\pm$ denote the band indices, {and $\mathbf{k}$ refers to the combination $(k,\gamma)$}. The vorticity is a non-Hermitian invariant and takes half integer values ($\mathbb{Z}/2$), because EP acts as a branch point on the complex-energy plane. Thus, it is essentially different from Hermitian topological invariants. 

Notably, it can be proved that EP can alter the winding number of
\begin{equation}
	w({\mathbf{k}}_{\text{EP}})=\frac{1}{2\pi i}\oint_{S^{1}_{\text{EP}}}d{k}\cdot\nabla_{\mathbf{k}}\log\det \mathcal{Q}(\mathbf{k}),\label{eq:EP-winding}
\end{equation}
by $w({\mathbf{k}}_{\text{EP}})=-2v_\pm$. The derivation details can be found in the  Appendix~\ref{winding_and_vorticity}. Thus, even though EP and its topological invariant are essentially non-Hermitain, it can determine Hermitian phase transition, and such a phase transition is different from conventional ones as shown in Fig.~\ref{fig:link}(a) by shifting energies.

Let us return to the 1D systems. By treating one of the two dimensions as the parameter to control the phase transition,  i.e., $\gamma$ in Fig.~\ref{fig:EP}(b),
the above topological invariant signifies the transition of topological phases in 1D. 
Because due to topological robustness, the circle $S^{1}_{\text{EP}}$ can be deformed to two loops, $L_1$
and $L_2$, as shown in Fig.~\ref{fig:EP}(b). As Eq.~(\ref{eq:EP-winding}) is essentially the same as the winding number of Eq.~(\ref{eq:chiral-winding}), the change of the winding number between $L_1$ and $L_2$
equals to the topological invariant of the EP. Thus, surprisingly, we find that non-Hermitian EPs can induce Hermitian topological phase transitions.

\vspace{0.3cm}

\section{Non-Hermitian braidings and Hermitian bulk-boundary correspondence}

\begin{figure}[t]
	\includegraphics[width=0.9\columnwidth]{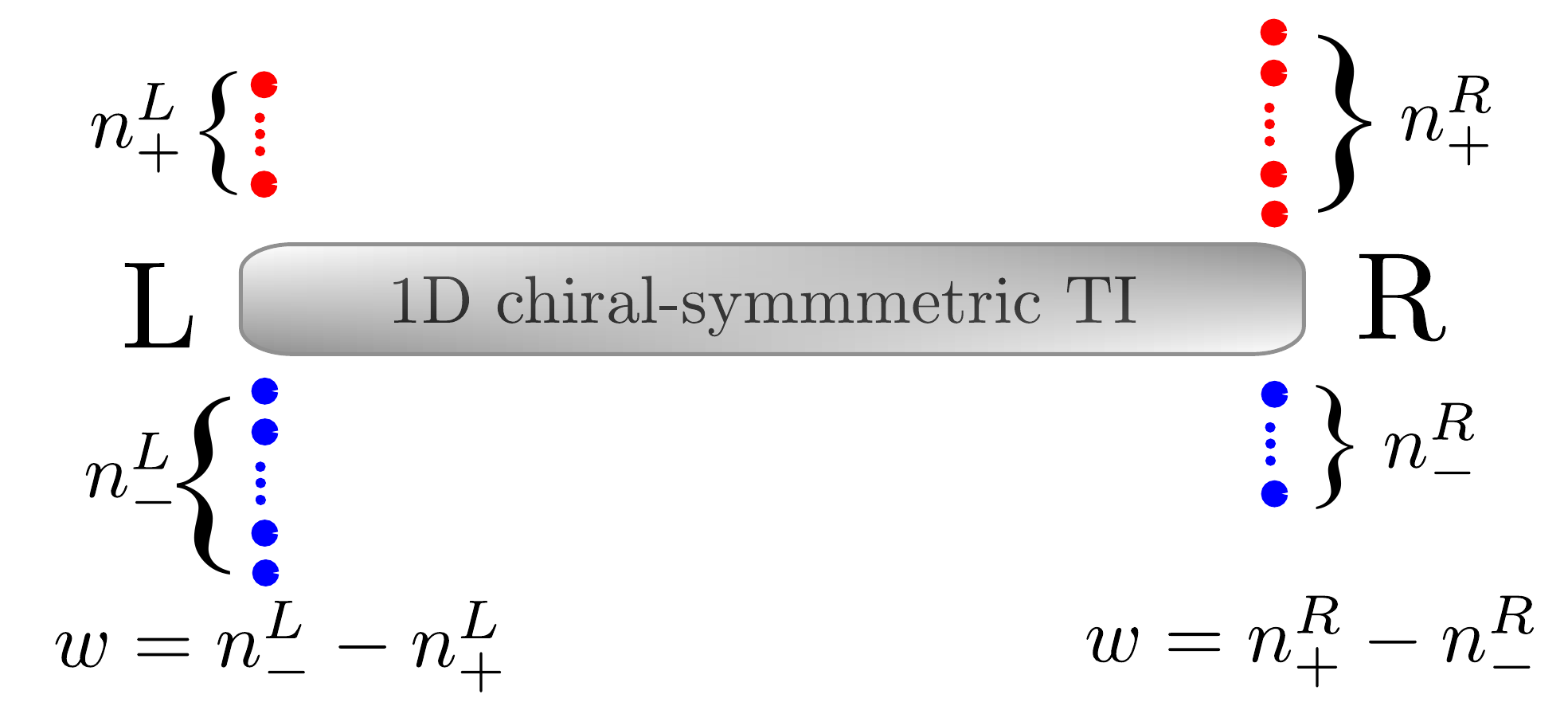}
	\caption{Hermitian bulk-boundary correspondence in 1D 
		chiral-symmetric topological insulators. The red or blue
		dots denote zero-energy modes localized at the left
		or right boundary.
		\label{fig:bbc}}
\end{figure}

According to Hermitian bulk-boundary 
correspondence, the non-trivial bulk topology corresponds to 
zero-energy topological boundary states, i.e., 
$w=n_{+}^{R(L)}-n_{-}^{R(L)}$ up to a sign factor. Here, $n_{+}^{R(L)}$ and $n_{-}^{R(L)}$ denote 
the number of zero-energy modes of $|\Psi_{+}\rangle$ and 
$|\Psi_{-}\rangle$ in Eq.~\eqref{eq:chiral-zero} that are exponentially localized at the right (left) boundary. Their relation with the bulk non-trivial winding number $w$ is illustrated in Fig.~\ref{fig:bbc}.


We have shown that the winding number comes from the non-Hermitian braidings using the linking number of Eq.~\eqref{eq:link}.
In one-dimensional non-Hermitian systems, this bulk winding number results in  non-Hermitian skin effect (NHSE)~\cite{kunstBiorthogonal2018,wang_skin_2018,Murakami_prl_2019,BorgniaPRL2020,Fang_winding_2020,Masatoshi_winding_2020}, manifested as the
localization of all wavefunctions on boundaries.
Note that this phenomenon has been observed recently in experiments~\cite{xiao_non-hermitian_2020,helbig_generalized_2020}. 
As the winding number comes from the linking number, we may regard that the NHSE corresponds to the bulk non-Hermitian braidings.

To further establish the relation between non-Hermitian braidings and Hermitian bulk-boundary correspondence, we show that the NHSE in $Q$ gives rise to topological boundary modes in $H$ under semi-infinite boundary conditions (SIBC), following~\cite{Masatoshi_winding_2020,Okuma-pseudospectrum}. These boundary conditions mean that the system extends to infinite with a single right (left) boundary. Let us consider a non-Hermitian 1D system $Q$ under SIBC. Under this boundary condition, all modes with non-trivial winding number in $Q$ exist and exhibit NHSE~\cite{Masatoshi_winding_2020}. Hence, as we consider the winding number to be non-trivial at the zero energy, i.e., $w\neq 0$ in Eqs.~\eqref{eq:chiral-winding} and \eqref{eq:link}, the zero-energy modes exist and are skin modes under SIBC. We note that the energy of any skin modes can be shifted to zero by adding a reference energy to the system.

The skin modes are exponentially localized at the right (left) boundary described by $\{E_{i,+},e^{\alpha j}|\psi_{i,+}\rangle\}$ for $Q^\dagger$ and $\{E_{i,-},e^{\alpha j}|\psi_{i,-}\rangle\}$ for $Q$.
Here, $i$ is the energy index, $j$ the site index, and $\alpha$ the decaying factor. 
We focus on the zero modes of $Q^{\dagger}$ and $Q$, whose numbers
are $n_{+}^{R(L)}$ and $n_{-}^{R(L)}$, respectively. Here, the superscript $R(L)$ denotes the right (left) boundary of the system under SIBC. In constructing $H$ by $Q$ using Eq.~\eqref{eq:chiral-hal}, owing to Eq.~\eqref{eq:Q-zero}, only these
zero-energy modes keep the same localization feature as in $Q$, while all other modes cannot retain in $H$~\cite{footnote}. Hence, under SIBC, the topological zero-energy boundary modes in $H$ originate from the zero-energy modes in NHSE of $Q$, which correspond to the bulk non-Hermitian braidings.  
 
It is noted that in a finite system, instead of SIBC, open boundary conditions (OBC) shall be considered. In this case, the zero-energy skin modes that exist under SIBC may not appear under OBC. However, the relation can still be established by using the concept of pseudospectrum in non-Hermitian systems, as explained in Ref.~\cite{Okuma-pseudospectrum}. {Specifically, when these zero-energy modes do not appear under OBC, there would be corresponding modes appearing in the pseudospectrum of $Q$, which are almost the eigenstates of the system, and they give rise to the topological modes in the Hermitian system.}

\vspace{0.3cm}
\section{Typical topological phases with non-Hermitian braidings}

\begin{figure}
	\includegraphics[width=0.9\columnwidth]{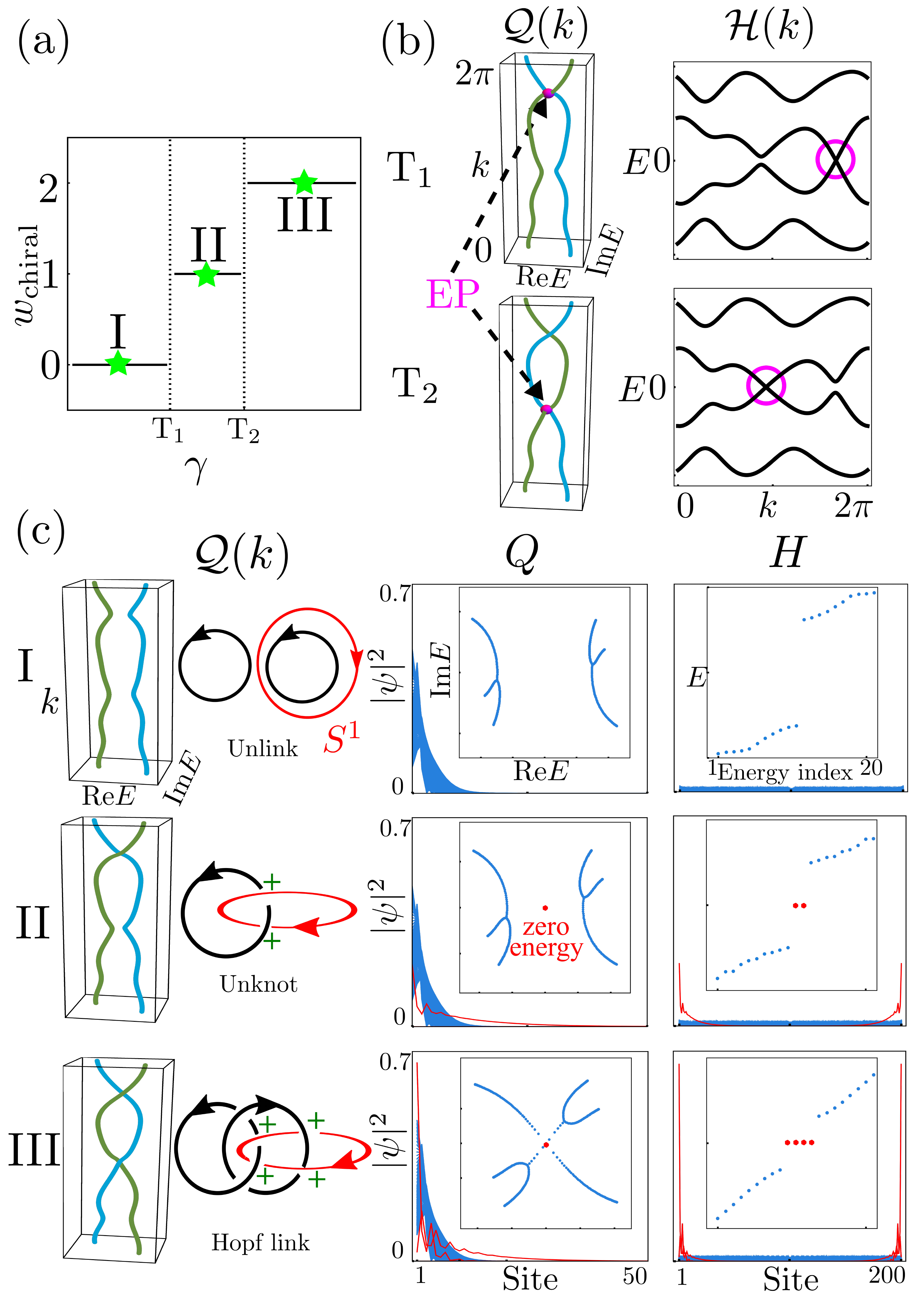}
	\caption{Typical two-band model. (a) The phase diagram for both $\mathcal{Q}(k)$
		and $\mathcal{H}(k)$ against $\gamma$. (b) The correspondence
		between EPs in non-Hermitian $\mathcal{Q}(k)$ and band crossings (open circle) in Hermitian $\mathcal{H}(k)$. 
		(c) The knots of non-Hermitian bands in the 1D BZ (left panel). The wavefunction profiles of $\mathcal{Q}(k)$ (middle panel) and $\mathcal{H}(k)$
		(right panel) with OBC. The insets in the middle panel shows
		the complex energy spectra of $\mathcal{Q}(k)$ with OBC. The insets
		in the right panel shows the energy spectra of $\mathcal{H}(k)$
		with OBC. {The red color refers to zero-energy modes determined by numerical calculations, which belong to the pseudospectrum of $\mathcal{Q}$.}
		The parameters are $\Delta=0.15,g=3\Delta,\kappa=1.4\Delta,c=1.5\Delta$.\label{fig:2band}}
\end{figure}

We now proceed to concrete models that demonstrate how the non-Hermitian braidings give rise to
Hermitian topological phases as well as their topological boundary states, whose phase transitions are mediated by critical points. Let us consider a two-band $\mathcal{Q}(k)$ with 
different non-Hermitian braidings, which has been realized recently~\cite{wangBraiding2021}. The
model Hamiltonian reads,
\begin{equation}
	\mathcal{Q}(k)=g\sigma_{1}+[\kappa\cos k+c\cos(2k)+i\Delta\sin(2k)-i\gamma]\sigma_{3}.\label{eq:model-q}
\end{equation}
The two complex-energy eigenvalues are 
$E(k)=\pm\left([\kappa\cos k+c\cos(2k)+i\Delta\sin(2k)-i\gamma]^{2}+g^{2}\right)^{1/2}$. As shown by I, II, and III in Fig.~\ref{fig:2band}(c) [three green stars in (a)], the braiding of two bands
in the $(\text{Re}E(k),\text{Im}E(k),k)$-space leads to different knots (black) in $\mathcal{Q}(k)$. 
These knots have different linking numbers with the zero-energy loop (red), and thus, they are different phases of $\mathcal{H}(k)$.
Specifically, I corresponds to
an unlinked phase having zero linking number; II is an unknot phase with linking number $+1$; III is
a Hopf link phase with linking number $+2$.

First, we show that the transformation of braiding structure mediated by EPs in $\mathcal{Q}(k)$,
as denoted by $T_{1}$ and $T_{2}$ in Fig.~\ref{fig:2band}(a),
leads to Hermitian topological phases transitions in $\mathcal{H}(k)$. Here,
$\mathcal{H}((k)= g\sigma_{1}\otimes\sigma_{1}+[\kappa\cos k+c\cos(2k)]\sigma_{1}\otimes\sigma_{3} +[\gamma-\Delta\sin(2k)]\sigma_{2}\otimes\sigma_{3}.$
We compare the energy spectra of $\mathcal{Q}(k)$ and $\mathcal{H}((k)$
at $T_{1}$ and $T_{2}$ in Fig.~\ref{fig:2band}(b). The position of band intersection
is determined by $E(k)=0$, as shown by the pink dots in (b).
At these points, $\mathcal{Q}(k)=g(\sigma_{1}\pm i\sigma_{3})$. It is doubly degenerate, but only has one eigenvector of $(\pm i,1)^{T}$, corresponding to order-2 EPs. Across these EPs, the winding number changes
by $+1$ due to the change of the non-Hermitian braiding. Thus, the Hermitian phase transitions are determined by these critical points.

Second, we turn to the relation between the non-Hermitian skin modes with zero
energy in $\mathcal{Q}$ and the topological boundary states
of Hermitian $\mathcal{H}$.
In the middle panel of Fig.~\ref{fig:2band}(c),
the wavefunction profiles of $\mathcal{Q}(k)$ under OBC for I, II, and III phases are plotted, which all possess NHSE. 
{
In particular, we numerically calculate the zero-energy states and their wavefunction profiles [red points/curves in (c)], which also exhibit skin effects localized at the boundary. However, numerical calculation alone cannot determine whether these are actual zero-energy modes or modes belonging to the pseudospectrum. This is because modes in the pseudospectrum could be very close to actual ones, as the pseudospectrum satisfies~\cite{Masatoshi_winding_2020,Okuma-pseudospectrum}
\begin{equation}
	\sigma_\varepsilon(Q)=\{E \in\mathbb{C} \mid ||(E-Q)|E \rangle || < \varepsilon \}, \label{pseudo-sp}
\end{equation}
where $E$ and $|E\rangle$ represent eigenvalue and eigenstate that are $\varepsilon$-close to the actual ones, and numerical calculations of the eigenvalues give only approximate results due to the underlying algorithm. In Appendix~\ref{zero-modes-psedudo}, we demonstrate that these modes belong to the pseudospectrum.} 
Furthermore, we note that in $Q$, the zero-energy modes themselves are skin modes, not topological boundary modes, as proved in Appendix~\ref{zero-modes-proof}.
In the right panel of (c),
the wavefunction profiles for the corresponding $\mathcal{H}(k)$ are plotted
under OBC. We can see that only II and III have
topological boundary states.
This is because for these two, there are zero-energy
skin modes in the pseudospectrum of $\mathcal{Q}$, while it is not the case for I. After constructing $H$ by $Q$, 
these zero-energy skin modes can give rise to the topological modes.


\vspace{0.3cm}
\section{Generalization to multi-band topological phases with higher-order EPs}

\begin{figure}
	\includegraphics[width=0.9\columnwidth]{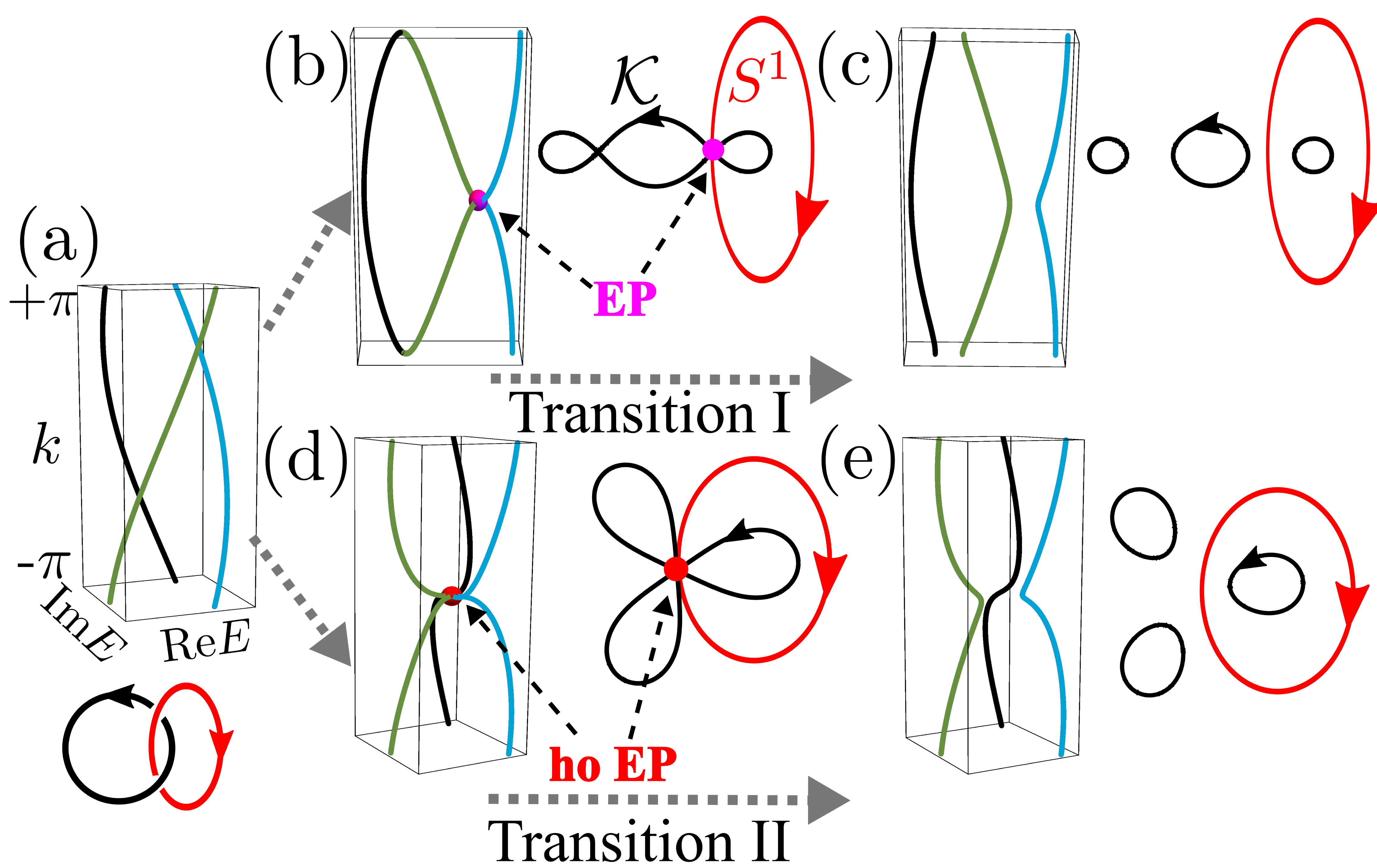}
	\caption{Generalization to multi-band models with higher-order EPs. Transition I of (a)$\rightarrow$(b)$\rightarrow$(c) is mediated by order-2 EPs. Here $\alpha=0$, and $\beta=0$, $1.89$, and $2.2$.
		Transition II of (a)$\rightarrow$(d)$\rightarrow$(e) is mediated by a higher-order EP (ho EP). Here $\beta=0$, and $\alpha=0$, $-1$, and $-1.1$. \label{fig:3band}} 
\end{figure}

The mechanism of non-Hermitian braidings can be generalized to 
multi-band models where higher-order EPs (ho EPs) may emerge. 
Here we discuss the case of triple-band $\mathcal{Q}(k)$, where
more than two bands could intersect at ho EPs. The non-Hermitian lattice model for $\mathcal{Q}(k)$ reads,
\begin{equation}\label{eq:tripe-band}
\mathcal{Q}(k)=\begin{pmatrix}
	0 & \beta & q(k)+\alpha\\
	1 & 0 & 0\\
	0 & 1 & 0
\end{pmatrix}.
\end{equation}
Here, $q(k)=\cos k+i \sin k$, and $\alpha$ and $\beta$ are model
parameters. With $\alpha=\beta=0$, the braiding of the three complex-energy bands forms a knot of circle, as shown in Fig.~\ref{fig:3band}(a). $\mathcal{H}(k)$ possesses a non-trivial phase if zero-energy loop links with this circle.

First, by increasing $\beta$, Transition I can be induced as shown by Figs.~\ref{fig:3band}(a,b,c). Similar to Fig.~\ref{fig:EP}, this transition is mediated by order-2 EPs, that is, the intersection of two bands. If one puts the zero-energy loop at such a EP [Fig.~\ref{fig:3band}(b)],
the change of knot topology from braiding results in a phase transition. Thus, in multi-band cases, order-2 EPs can still lead to 
topological phase transitions in $\mathcal{H}(k)$. 

Second, by decreasing $\alpha$, Transition II can be induced as shown by Figs.~\ref{fig:3band}(a,d,e). This transition is mediated by ho EP [red point in (d)]. It is different from Transition I, as there are three bands intersecting at this point. However, ho EP may also mediate topological phase transition. Because by putting zero-energy loop at this point, Transition II also signifies the change of the linking number,
resulting the phase transition in $\mathcal{H}(k)$. Therefore, similar to order-2 EPs, ho EP can mediate
the change of braiding structure, resulting in the topological phase transitions in $\mathcal{H}(k)$.

\section{DISCUSSION}


We have demonstrated that all the fundamental concepts of chiral-symmetric Hermtian topological phases, including bulk topology, topological phase transition, and bulk-boundary correspondence, have non-Hermitian origins. This non-Hermitian picture provides
a deeper understanding which is not possible by Hermitian theories, as signified by the discovery of a novel class of topological phase transitions characterized by the critical points. Hence, our work shows that non-Hermitian topology may be regarded as building blocks of Hermitian topological phases. {In view of the rich concepts in non-Hermitian physics, it would be worthwhile to explore further, e.g., in the following direction. In the case of  braiding structures formed by more than two energy bands, similar to the Hermitian systems~\cite{BouhonPRB2020,NurPRL2020,bouhonarxiv2022}, multi-gap(band) conditions may be taken into account. It is expected that partitioning energy gaps/bands into different groups could lead to different topologies in non-Hermitian $Q$, which may result in different Hermitian topologies of $H$. 


\section{ACKNOWLEDGMENTS}
{The authors are grateful to the anonymous referees for their valuable help in this work.} This work was supported by the Guangdong-Hong Kong Joint Laboratory of Quantum Matter, the NSFC/RGC JRS grant (N\_HKU774/21), and the CRF of Hong Kong (C6009-20G). W.B.R. was supported by the RGC Postdoctoral Fellowship (PDFS2223-7S05).

\appendix

\section{The relation between vorticity and winding number \label{winding_and_vorticity}}

	In this section, we prove that the topological invariant of order-2 EP, i.e., the vorticity, can determine the change of the winding number. 
	In 2D space, the vorticity of a EP reads,
	\begin{equation}\label{vorticity}
		v_\pm=-\frac{1}{2\pi}\oint_{S_\text{EP}^1}dk\cdot\nabla_{k\ } \arg\left[E_+\left(k\right){-E}_-\left(k\right)\right].
	\end{equation}
Here, $k$ refers to the momentum in the 2D space. The winding number around a EP is
	\begin{equation}
		w({k}_{\text{EP}})=\frac{1}{2\pi i}\oint_{S^{1}_{\text{EP}}}d{k}\cdot\nabla_{k}\log\det \mathcal{Q}(k).\label{eq:EP-winding-ap}
	\end{equation}
	The vorticity is a non-Hermitian invariant and takes half integer values ($\mathbb{Z}/2$), because EP acts as a branch point on the complex-energy plane. In the meantime, the winding number takes integer values. 
	
	For the relevant two bands around an order-2 exceptional point (suppose it is located at $k=0$) of $Q(k)$, the effective Hamiltonian can be written as
	\begin{equation}
		Q_{eff}(k)=d_1(k)\sigma_1+d_2(k)\sigma_2+d_3(k)\sigma_3,
	\end{equation}
	where $\sigma_i$'s (i=1,2,3) are Pauli matrices. The energy eigenvalues are
	\begin{equation}
		E_\pm(k)=\pm d(k)=\pm \sqrt{d_1(k)^2+d_2(k)^2+d_3(k)^2}.
	\end{equation}
	The winding number of Eq.~\eqref{eq:EP-winding-ap} can be calculated as
	\begin{equation}\label{winding}
		\begin{aligned}
			w({k}_{\text{EP}})& = \frac{1}{2\pi i} \oint_{S_{EP}^1} dk\cdot\nabla_k \log\det Q_{eff}(k)\\
			& = \frac{1}{2\pi i} \oint_{S_{EP}^1} dk\cdot\nabla_k \log E_+(k)E_-(k)\\
			& = \frac{1}{2\pi i} \oint_{S_{EP}^1} dk\cdot\nabla_k (\log E_+(k)+ \log E_-(k))\\
			& = \frac{1}{2\pi i} \oint_{S_{EP}^1} dk\cdot\nabla_k (\log d(k)+ \log (-d(k))).
		\end{aligned}
	\end{equation}
	The two vorticities ($v_\pm$ and $v_\mp$) according to Ref. [26] can be calculated as
	\begin{equation}\label{v1}
		\begin{aligned}
			v_\pm& = -\frac{1}{2\pi } \oint_{S_{EP}^1} dk\cdot\nabla_k \arg [E_+(k)-E_-(k)]\\
			& =  -\frac{1}{2\pi } \oint_{S_{EP}^1} dk\cdot\nabla_k \arg [2d(k)]\\
			& =  -\frac{1}{2\pi} \oint_{S_{EP}^1} dk\cdot\nabla_k \left(-i\log\frac{2d(k)}{|2d(k)|} \right)\\
			& = -\frac{1}{2\pi i} \oint_{S_{EP}^1} dk\cdot\nabla_k \log d(k).
		\end{aligned}
	\end{equation}
	Note we use the facts that $\arg \left(z\right)=-i\log\left(\frac{z}{\left|z\right|}\right)$  in the third step and that $\left|\log d\left(k\right)\right| $ is a single-valued function in the last step. And similarly,
	\begin{equation}\label{v2}
		\begin{aligned}
			v_\mp& = -\frac{1}{2\pi } \oint_{S_{EP}^1} dk\cdot\nabla_k \arg [E_-(k)-E_+(k)]\\
			& =  -\frac{1}{2\pi } \oint_{S_{EP}^1} dk\cdot\nabla_k \arg [-2d(k)]\\
			& = -\frac{1}{2\pi i} \oint_{S_{EP}^1} dk\cdot\nabla_k \log (-d(k)).
		\end{aligned}
	\end{equation}
	In view of the winding number $w(k_\text{EP})$ of Eq.~\eqref{winding} and the vorticities ($v_\pm$ and $v_\mp$) of Eqs.~\eqref{v1} and \eqref{v2}, it is clear that
	\begin{equation}
		w({k}_{\text{EP}})=-\left(v_\pm+v_\mp\right)
	\end{equation}
	We note that $v_\pm$=$v_\mp$ because
	\begin{equation}
		\begin{aligned}
			v_\mp& = -\frac{1}{2\pi i} \oint_{S_{EP}^1} dk\cdot\nabla_k \arg[-2d(k)]\\
			& =-\frac{1}{2\pi i} \oint_{S_{EP}^1} dk\cdot\nabla_k (\arg[2d(k)]+\pi)\\
			& = -\frac{1}{2\pi i} \oint_{S_{EP}^1} dk\cdot\nabla_k (\arg[2d(k)])\\
			&=v_\pm
		\end{aligned}
	\end{equation}
	Thus, we have
	\begin{equation}
		w({k}_{\text{EP}})=-2v_\pm\ ({\text{or }-2v}_\mp).
	\end{equation}
	Therefore, the winding number of $w({k}_{\text{EP}})$ is determined by the vorticity of exceptional point. While the vorticities take half integer values, the winding number takes integer values. This is different from the ordinary phase transition by shifting energies.

{

\section{The zero-energy modes in Fig. 4(c) belong to pseudospectrum \label{zero-modes-psedudo}}
In this section, we demonstrate that the zero-energy modes in Fig.~4(c) II, obtained numerically by diagonalizing
the Hamiltonian, belong to the pseudospectrum.  

\begin{figure*}
	\centering
	\includegraphics[width=0.8\textwidth]{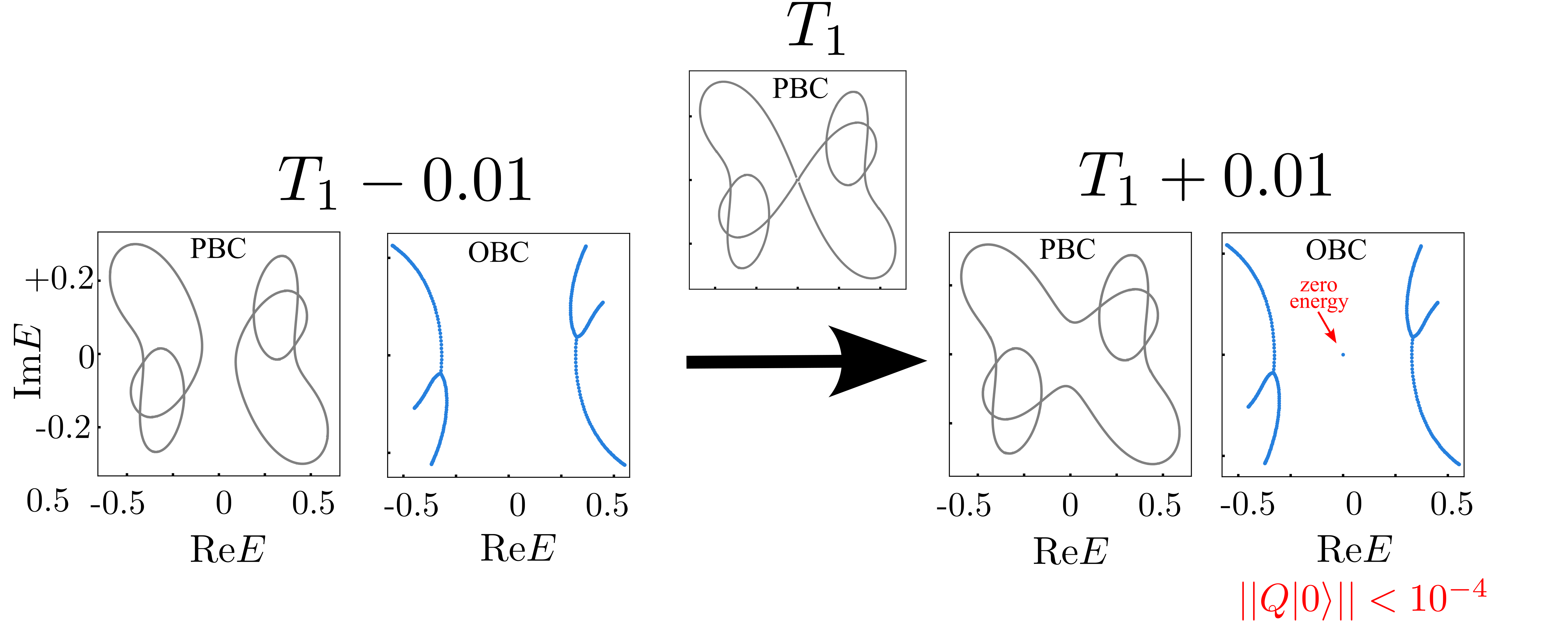}
	\caption{The evolution of the PBC and OBC spectra around the phase transition point at $T_1$. Numerical calculation shows that the zero modes appear immediately after $T_1$, determined by the criteria $||Q|0\rangle||<10^{-4}$.
		\label{fig:AroundT1}}
\end{figure*}

First, we investigate the behaviour around the phase transition point $T_1$ in Fig.~\ref{fig:2band}. As shown in Fig.~\ref{fig:AroundT1}, we calculate the spectra immediately before $T_1$, i.e., $T_1-0.01$, and immediately after $T_1$, i.e., $T_1+0.01$. We find that, there exists a zero mode $|0\rangle$ satisfying, 
\begin{equation}\label{pseudo-zero}
	||Q|0\rangle||<10^{-4},
\end{equation}
at $T_1+0.01$, while there is no such mode at $T_1-0.01$. Note that numerical calculations give only approximate
values, which means this mode may also belong to pseudospectrum according to Eq.~\eqref{pseudo-sp}. To determine whether this is an actual zero mode of $Q$, we note that the phase transition is tuned by a continuous parameter $\gamma$, and thus, the spectrum shall change continuously. Hence, if zero modes exist after $T_1$, some modes shall approach zero before $T_1$. However, there are no such modes at $T_1-0.01$. Thus, by the continuous property of the Hamiltonian, this isolated zero mode shall not be an actual mode of $Q$.

Second, having ruled out the zero modes of Eq.~\eqref{pseudo-zero} as actual zero modes of $Q$, we proceed to demonstrate that this zero mode belongs to the pseudospectrum. According to Refs.~\cite{Masatoshi_winding_2020,Okuma-pseudospectrum}, the pseudospectrum satisfies
\begin{equation}
	\lim_{\epsilon\rightarrow 0} \lim_{N\rightarrow \infty} \sigma_\epsilon(Q_\text{OBC})=\sigma(Q_\text{SIBC}),
\end{equation}
where $N$ is the system size, and $Q_\text{OBC}$ and $ Q_\text{SIBC}$ are the Hamiltonians of OBC and SIBC systems, respectively. The above equation means that after taking the limits, the pseudospectrum $ \sigma_\epsilon(Q_\text{OBC})$ is equal to the SIBC spectrum $ \sigma(Q_\text{SIBC})$, which is given by the union of $ \sigma(Q_\text{PBC})$ and the whole area of $E \in \mathbb{C}$ enclosed by $ \sigma(Q_\text{PBC})$ with a non-zero spectral winding number. Here, $\sigma(Q_\text{PBC})$ refers to the PBC spectrum. Regarding the region immediately after $T_1$, 
as shown in Fig.~\ref{fig:AroundT1}, for a fixed $\varepsilon$ with a large $N$, the pseudospectrum is given by the region enclosed by the PBC spectrum with non-zero winding number, which includes the zero energy point and differs from the region immediately before $T_1$. Thus, after $T_1$, the zero mode can appear in this pseudospectrum region, which is determined by $\varepsilon=10^{-4}$ of Eq.~\eqref{pseudo-zero} in numerical calculations.
}

\section{The zero-energy modes in Fig. 4(c) are not topological modes in $Q$ \label{zero-modes-proof}}

In this section, we further demonstrate that the zero-energy modes in Fig. 4(c) II and III in the main text, are skin modes, not topological modes of $Q$. To do this, we first show that these modes have the common feature of skin modes, and then show that other than non-Hermitian braidings, there is no additional non-trivial topological structure associated with the model Hamiltonian of Eq. (9) in the main text.

First, the skin modes have the common feature of appearing inside the periodic boundary spectrum. This is because they have a non-trivial winding number, which requires to be enclosed by the periodic boundary spectrum. In Fig.~\ref{fig:spectrum}, we plot the spectra of $Q\left(k\right)$ for I, II and III under periodic boundary condition (PBC), corresponding to Fig. 4(c) in the main text, and then compare them with those under open boundary condition (OBC). We can see that the OBC spectra (blue) all fall inside the PBC spectrum (gray) for I, II and III. In particular, for the zero modes of OBC in II and III, as highlighted by red arrows, they also fall inside the PBC spectrum (Gray). Thus, these zero modes has the feature of skin modes.

\begin{figure}
	\centering
	\includegraphics[width=0.99\columnwidth]{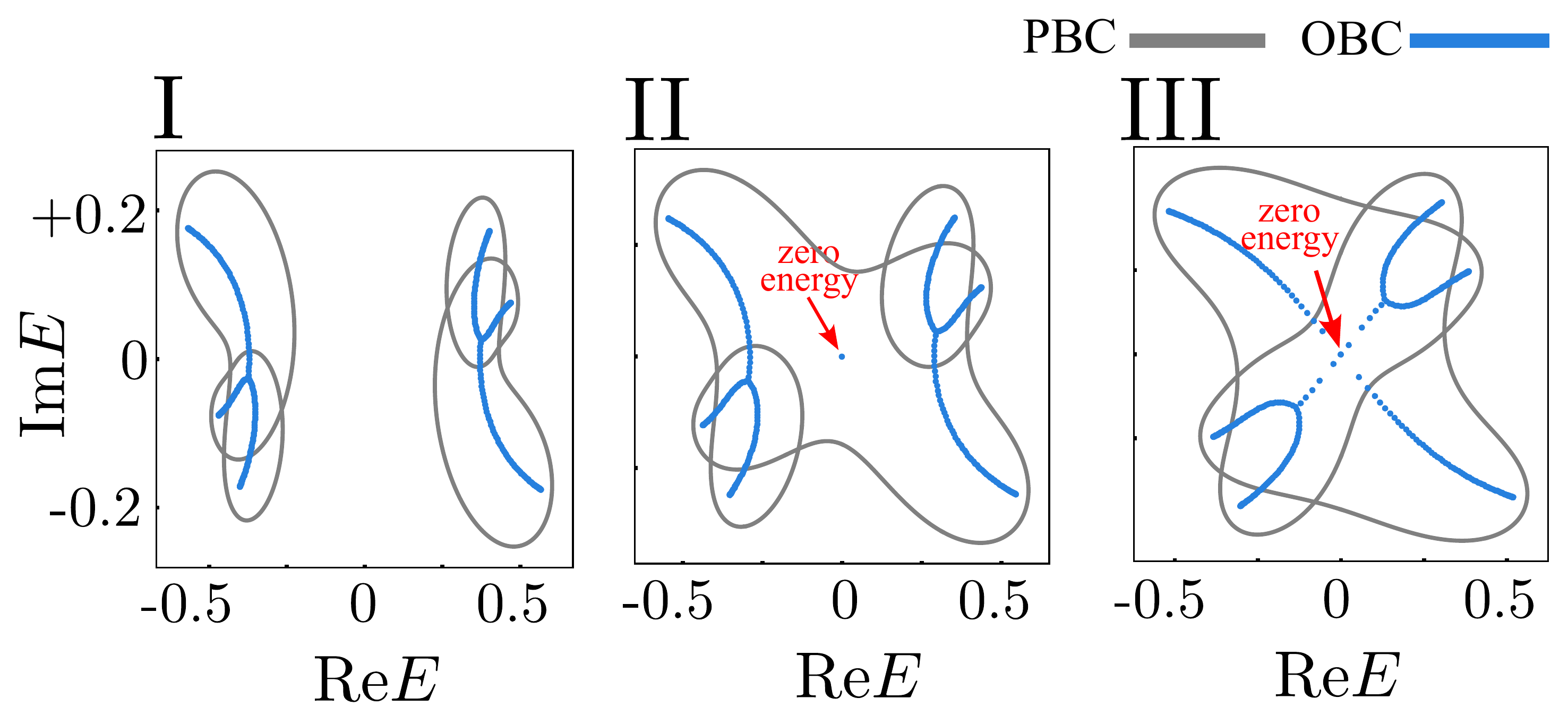}
	\caption{The comparison between spectra under periodic boundary condition (PBC, gray) and open boundary condition (OBC, blue). The zero modes in OBC spectra are highlighted by red arrows, which fall inside the PBC spectra.
		\label{fig:spectrum}}
\end{figure}

\begin{figure}
	\centering
	\includegraphics[width=0.99\columnwidth]{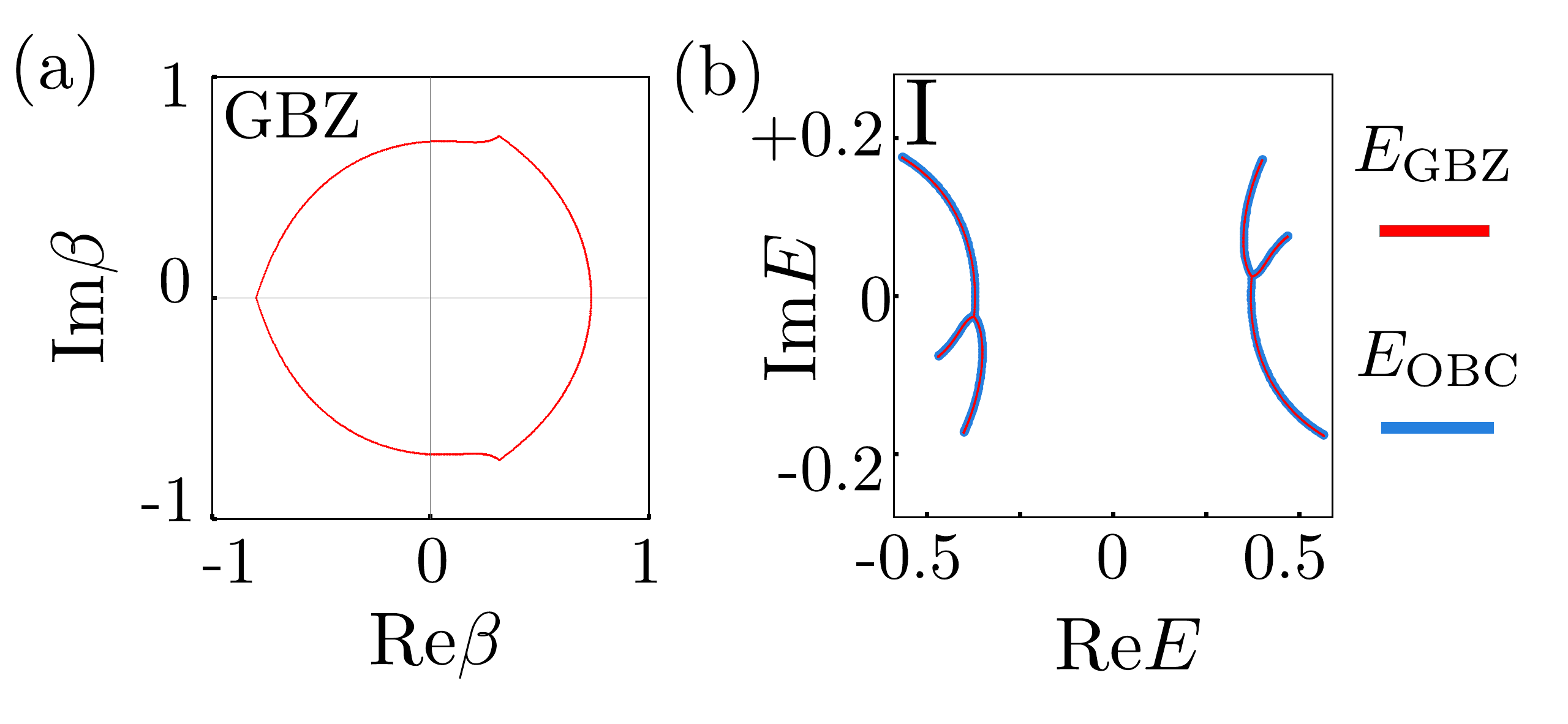}
	\caption{(a) Generalized Brillouin zone (GBZ) calculated for Fig. 4(c) I in the main text. (b) The comparison between the GBZ spectrum $E_{\text{GBZ}}$ and the OBC spectrum $E_{\text{OBC}}$.
		\label{fig:GBZ}}
\end{figure}

Second, the above criteria, however, cannot fully rule out the possibility that these zero-energy modes are topological modes. Because there is a special chiral symmetry $\sigma_y$ for $Q\left(k\right)$ of Eq. (8) in the main text, to achieve the goal of ruling out topological modes, we need to calculate the winding number (denoted as $w_2$) associated with this special chiral symmetry $\sigma_y$. A non-trivial $w_2$ indicates the existence of topological zero modes, while a trivial one cannot. However, since $Q\left(k\right)$ is non-Hermitian, we cannot directly calculate $w_2$ using the Bloch Hamiltonian, because of the difference between PBC and OBC spectrum. Following Refs. [46,47] in the main text, we use the non-Bloch band theory and calculate the generalized Brillouin zone (GBZ) first, and then use the GBZ to compute the non-Bloch topological invariant $w_2$.

\begin{figure}
	\centering
	\includegraphics[width=0.7\columnwidth]{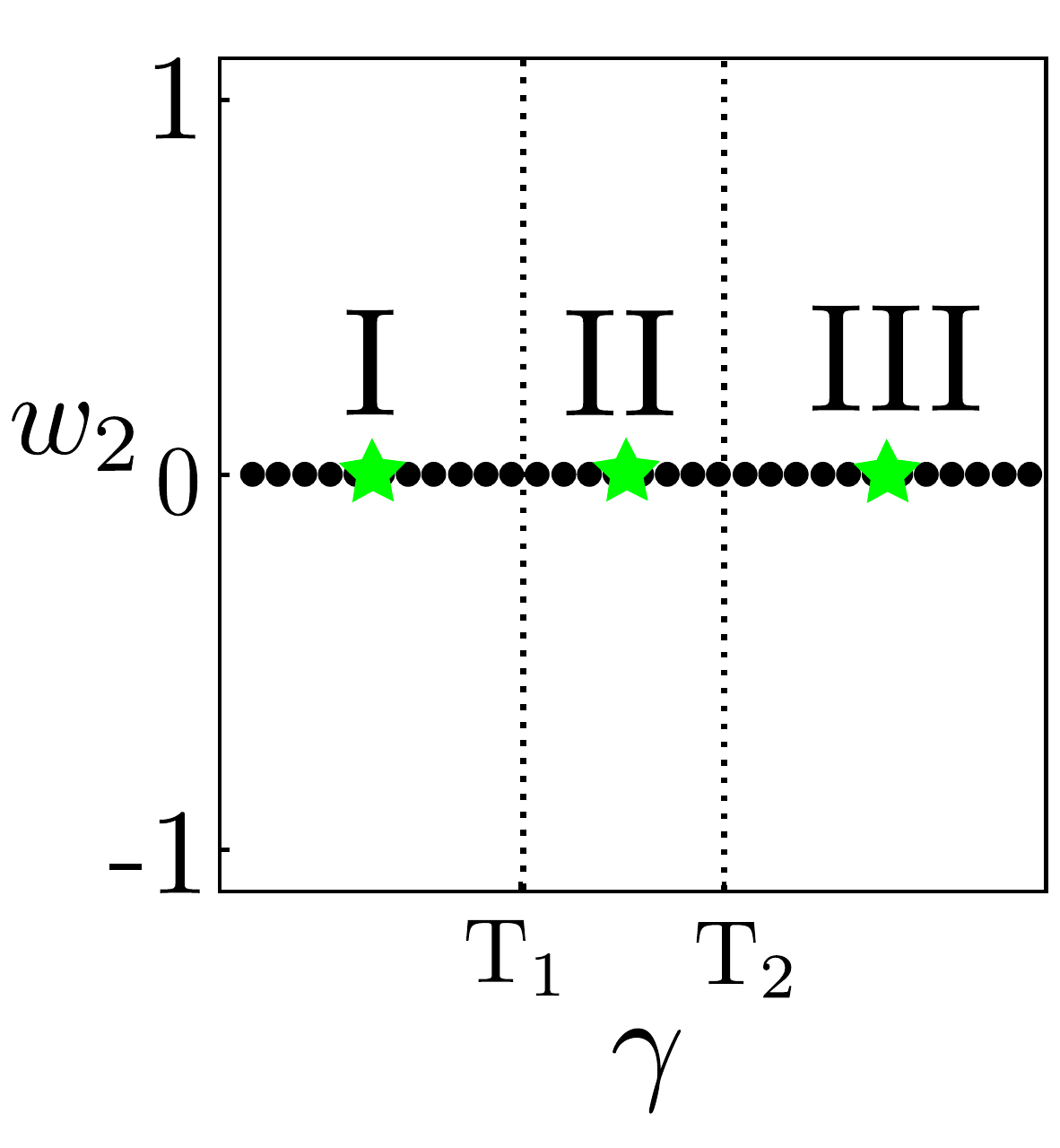}
	\caption{The non-Bloch winding number calculated by the generalized Brillouin zone (GBZ). It corresponds to the special chiral symmetry of the non-Hermitian $Q\left(k\right)$. The plot range is the same as Fig. 4(a) in the main text.
		\label{fig:phase-transition}}
\end{figure}

In Fig.~\ref{fig:GBZ}(a), the GBZ for Fig. 4(c) I in the main text is obtained, which correspond to Fig.~\ref{fig:spectrum} I above. We use this GBZ to compute the band spectrum $E_\text{GBZ}$ and plot it in Fig.~\ref{fig:GBZ}(b). Comparing with OBC spectrum $E_\text{OBC}$, we can see that GBZ can well predict the behavior of open boundary systems. 

We proceed to compute the non-Bloch winding number $w_2$ based on GBZ. First, we perform a unitary transformation $U=\exp\left(i\frac{\pi}{4}\sigma_x\right)$ to obtain the standard off-diagonal form of $Q(k)$,
\begin{align}
	UQ\left(k\right)U^{-1}=
	\begin{pmatrix}
		0&R_+\left(\beta\right)\\R_-\left(\beta\right)&0\\
	\end{pmatrix},  
\end{align}
where the chiral symmetry operator becomes $U\sigma_yU^{-1}=-\sigma_z$. Note that we have substituted $k$ with $\beta$ for GBZ ($k\rightarrow \beta=e^{k}$). Then according to Eq. (8) of Ref. [47] in the main text, the winding number $w_2$ can be computed as,
\begin{align}
	w_2=-\frac{w_+-w_-}{2},{\ \ \ \ w}_\pm=\frac{1}{2\pi}\left[\arg\ R_\pm\left(\beta\right)\right]_{C_\beta},                
\end{align}
where $\left[\arg\ R_\pm\left(\beta\right)\right]_{C_\beta}$ means the change of phase of $R_\pm\left(\beta\right)$ as $\beta$ goes along the generalized Brillouin zone $C_\beta$ in a counterclockwise way.

As shown in the Fig.~\ref{fig:phase-transition} above, we have calculated the non-Bloch winding number $w_2$ numerically. We find that it is trivial, i.e., $w_2=0$, for all $\gamma$ in the range same as Fig. 4(a) in the main text. In particular, for the parameters of I, II and III in Fig. 4(c) (or Fig.~\ref{fig:spectrum} above), their winding numbers are all trivial ($w_2=0$). Therefore, we can conclude that the zero modes in II and III are not topological modes. In other words, they are skin modes. 


{

\section{An example outside class AIII\label{extra-example}}
In the main text, we consider $H$ in Hermitian class AIII constructed by $Q$ using Eq.~(\ref{eq:chiral-hal}). In this section, we provide an example outside class AIII, namely, a model in class BDI.

The momentum-space Hamiltonian reads, 
\begin{equation}
Q(k)=g\sigma_1+\left[c\sin{\left(2k\right)}+ic\cos{\left(2k\right)}-i\gamma\right]\sigma_3,
\end{equation}
where $g$, $c$, and $\gamma$ are model parameters, and $\sigma_i$'s are standard Pauli matrix. Here, $Q(k)$ has a similar braiding topology as that in Fig.~\ref{fig:2band}(c)-III, for parameters $c=0.15,g=0.2,$ and $\gamma=0.2$.

Based on  $Q(k)$, we can construct the Hermitian Hamiltonian, 
\begin{equation}
	H\left(k\right)=g\tau_1\otimes\sigma_1+c\sin{\left(2k\right)}\tau_1\otimes\sigma_3-\left[c\cos{\left(2k\right)}-\gamma\right]\tau_2\otimes\sigma_3.
\end{equation}
Besides chiral symmetry, this Hamiltonian has the the following symmetries, \\
(i) charge-conjugation symmetry:
\begin{equation}
	CH\left(k\right)C^{-1}=-H\left(-k\right),\qquad C=\tau_0\otimes\sigma_3\mathcal{K},
\end{equation}
(ii) time-reversal symmetry:
\begin{equation}
	TH\left(k\right)T^{-1}=H\left(-k\right),\qquad T=\tau_0\otimes\sigma_1\mathcal{K}.
\end{equation}
Here $\mathcal{K}$ is the complex conjugation operation. Because $C^2=T^2=+1$, the model Hamiltonian belongs to class BDI. It is straightforward to verify that $H\left(k\right)$ has a non-trivial topology due to the braiding in $Q(k)$.

}

\bibliographystyle{apsrev4-2}
\bibliography{Chiral-non-hermi}

\begin{thebibliography}{72}
\bibitem [{foo()}]{footnote}%
  \BibitemOpen
  \href@noop {} {}\bibinfo {note} {Away from zero energy, the eigenequation can
  also be obtained by the non-Hermitian components $Q$ and $Q^{\dagger}$, i.e.,
  $Q\psi_{-}=E\psi_{+}$ and $Q^{\dagger}\psi_{+}=E\psi_{-}$, which yield
  $QQ^{\dagger}\psi_{+}=E^{2}\psi_{+},\quad
  Q^{\dagger}Q\psi_{-}=E^{2}\psi_{-}.$ However, this becomes a coupled
  eigenvalue problem, and the information from either $Q$ or $Q^{\dagger}$
  alone cannot be retained in the Hermitian Hamiltonian $H$.}\BibitemShut
  {Stop}%
\bibitem [{poi()}]{pointgap}%
  \BibitemOpen
  \href@noop {} {}\bibinfo {note} {{ Given non-Hermitian $Q_1$ and $Q_2$, they
  have the same point-gap topology about a refence point $E_P$ on the
  complex-energy plane, if their spectra can be smoothly deformed to each other
  without crossing $E_P$. }}\BibitemShut {NoStop}%
\bibitem [{cal()}]{calnote}%
  \BibitemOpen
  \href@noop {} {}\bibinfo {note} {{Here, the simplification refers to the
  comparison between integration and counting crossings. There might be another
  computational cost arising in calculating the linking number. That is, to
  obtain the linking number, it is required to identify all the eigenvalues,
  while in Eq.~\eqref{eq:chiral-winding} only the computation of the
  determinant is required.}}\BibitemShut {Stop}%
\end{thebibliography}

\end{document}